\documentstyle[psfig,epsf,conf-X]{article}
\begin{document} 
\small
\heading{%
%Begin Heading
%
Clustering in the X-ray Universe
% End Heading
}
\par\medskip\noindent
\author{%
%Begin Author names
X. Barcons$^{1}$, F.J. Carrera$^{2,1}$, M.T. Ceballos$^{1}$, S. Mateos$^{1,3}$
%End Author names
}
\address{%
%First address
Instituto de F\'\i sica de Cantabria (CSIC-UC), 39005 Santander, Spain
}
\address{%
% Second Address
Mullard Space Science Laboratory, University College London, UK
}
\address{%
% Third Address
Departamento de F\'\i sica Moderna, Universidad de Cantabria, 39005 Santander, Spain
}

\begin{abstract}
In this paper we discuss various possibilities of using X-ray
observations to gain information about the large-scale structure of
the Universe. After reviewing briefly the current status of these
investigations we explore different ways of making progress in this
field, using deep surveys, large area surveys and X-ray background
observations. 

\end{abstract}
\section{Introduction}

X-ray emission in the Universe arises in intense gravity
environments.  At high galactic latitudes, active galactic nuclei
(AGN) and other emission line galaxies dominate the source counts at
all explored fluxes, with galaxy clusters being the second most
abundant source class.  

Recent {\it ROSAT} deep surveys (\cite{b01}, \cite{b02}, \cite{b03})
have shown that most of the soft X-ray volume emissivity in the
Universe arises at redshifts $z>1-2$ (\cite{b04}). The AGN and star
formation rate per unit volume follow a remarkably similar evolution
rate in the Universe (\cite{b05}) and therefore they can both be used
as tracers of the evolution of large-scale structure in the
Universe. Using AGN and star-forming galaxies as tracers of cosmic
inhomogeneities is most sensitive to intermediate redshifts ($z\sim
2$), providing a critical link between cosmic microwave background
studies (which map the $z\sim 1000$ Universe) and local galaxy surveys
($z\sim 0$).

In this paper we briefly review the current status of the use of X-ray
observations towards the study of large-scale structure. More details
are presented in \cite{b29}. Then we explore possibilities
of making qualitative progress in this field by carrying out different
types of X-ray surveys.

\section{What we know so far}

At galactic latitudes $\mid b\mid > 20^{\circ}$ the contribution from
the Galaxy to the X-ray sky is small: less than 10\% of the X-ray
background above 2 keV is due to galactic emission, absorption is
negligible above this photon energy and a census of X-ray sources down
to any flux limit exhibits less than 10-20\% of galactic stars.
Observations of the X-ray background at high galactic latitudes and
photon energies above 2 keV can therefore be used to map the
extragalactic X-ray sky.  

\subsection {The isotropy of the X-ray background} 

The all-sky distribution of the X-ray background for cosmological
purposes has been best mapped by the HEAO-1 mission.  A galactic
anisotropy dominates the large-scale anisotropy, but this can be
modelled out (\cite{b06}). A dipole contribution is detected in the
X-ray sky, in rough alignment with the direction of our motion with
respect to the Cosmic Microwave Background frame (\cite{b07},
\cite{b08}). The amplitude of this dipole accounts for both the
kinematical effect of our motion (the Compton-Getting effect) and the
excess X-ray emissivity associated with the structures which are
pulling us.  These two effects are expected to be of the same order
(\cite{b09}) and the analysis done in \cite{b08} shows this to be the
case.  However, in an analysis of the ROSAT all-sky data at lower
photon energies (which have the disadvantage of a larger contamination
from the Galaxy) Plionis \& Georgantopoulos \cite{b10} find a dipole
several times larger than the expected kinematical dipole.  The
difference between both results might be partly affected by the
elimination of X-ray bright clusters in the Scharf et al analysis, as
clusters are known to be a largely biased population (\cite{b11}). The
bias parameter derived from the XRB dipole is large ($b_X\sim 3-6$).

Treyer et al \cite{b12} have analyzed higher order multipoles of the
HEAO-1 A2 X-ray background.  The discrete nature of the XRB
contributes a constant term to all multipoles which scales as $\propto
S_{cut}^{0.5}$, where $S_{cut}$ is the minimum flux at which sources
have been excised. Treyer et al detect a signal growing towards
lower-order multipoles which is consistent with a gravitational
collapse picture, as predicted by \cite{b09}.  Excluding the
dipole, this analysis yields a moderate bias parameter for the X-ray
sources ($b_X\sim 1-2$).

On smaller (a few degrees) angular scales, probing linear scales of
hundreds of Mpc, the `excess fluctuations' technique has been used
often in the analysis of the XRB.  The way this works is by modelling
the distribution of XRB intensities on a given angular scale in terms
of confusion noise, plus a contribution coming from source clustering
(\cite{b13}). These studies have yielded so far only upper limits
for the excess fluctuations: $< 2\%$ on scales of $5^{\circ}\times
5^{\circ}$ (\cite{b14}) and $<4\%$ on scales
$1^{\circ}\times 2^{\circ}$ (\cite{b15}).  We discuss later
what is the expected signal and how it could be measured.

Yet on smaller (a few arcmin) angular scales, which probe the
galaxy-galaxy clustering scale, data from X-ray imaging telescopes has
been used. The autocorrelation function of the XRB on these scales
should reflect the clustering of high redshift X-ray sources in the
nonlinear regime (\cite{b16}).  Soltan et al \cite{b17} have found a
strong positive detection for angular separations 0.3-20$^\circ$ which
is, however, difficult to interpret as both the Galaxy and the Local
Supercluster could contribute to this.

\subsection {Clustering of X-ray selected AGN}

Studying the clustering of X-ray selected AGN is likely to be the most
direct way to map the structure of the X-ray sky. At soft X-ray
energies this requires fairly deep surveys (going below $\sim
10^{-14}\, {\rm erg}\, {\rm cm}^{-2}\, {\rm s}^{-1}$) as otherwise
very few objects at $z>1$, where most of the X-ray volume
emissivity is produced, would be sampled.  

Carrera et al \cite{b18} have analysed a set of `pencil beam' medium and
deep ROSAT images containing 200 X-ray selected AGN, sampling a
redshift interval $z\sim 0-2$.  The net result is the detection of
X-ray selected AGN clustering which is relatively weak (the 3D
correlation length is $r_0< 5\, h^{-1}\, {\rm Mpc}$, for $h=H_0/(100\,
{\rm km}\, {\rm s}^{-1}\, {\rm Mpc}^{-1})$) and strongly evolving with
redshift (faster than comoving). At much brighter flux limits \cite{b19} used the ROSAT All Sky Survey Sources to derive a 2D
correlation function that, when translated to 3D with an appropriate
catalogue depth, yields a larger correlation length ($r_0\sim 6\,
h^{-1}\, {\rm Mpc}$).

\subsection {Do X-rays trace mass?}

In \cite{b29} we compile various measurements of the bias
parameter for X-ray sources and in particular for X-ray selected AGN
and the XRB.  The bias parameter
is likely to be redshift dependent.  For a simple model where all
objects form at the same early redshift, Fry \cite{b20} finds
$b_X(z)=b_X(0)+z(b_X(0)-1)$, which implies that at high z the bias 
parameter could be large.

The other effect that comes into play, especially when using the XRB,
is that at low redshift clusters become more numerous and their
imprint in the local XRB features becomes more important.  As clusters
are a strongly biased source population (\cite{b11}
estimate $b_X(0)\sim 4$) it is not surprising that the amplitude of
the XRB dipole calls for large bias factors, but higher order
multipoles (sensitive to more distant sources) do not.

Within present knowledge and uncertainties, the bias factor for the
AGN as the dominant X-ray source population, appears to take moderate
values $b_X=1-2$ at low to intermediate redshifts. Indeed at higher
redshifts the AGN population might be more strongly biased.

\section{Deep  hard X-ray surveys}

Deep surveys, particularly at hard photon energies, are a key
ingredient to forthcoming studies af the large-scale structure of the
X-ray Universe.  Currently popular models for the X-ray background
assume a population of AGN with a distribution of absorbing columns
(\cite{b21}, \cite{b22}, \cite{b23}), where
most of the X-ray energy produced by accreting black holes is absorbed
and re-radiated in the infrared (\cite{b24}, \cite{b25}).  Several claims have been made that the absorbed AGN population
evolves differently than the unabsorbed one (\cite{b26},
\cite{b30}, \cite{b31}).  As most of the energy content in
the XRB resides at 30 keV, it is crucial to explore harder photon
energies than previously achieved with the ROSAT deep surveys.

\begin{figure}[!h]
\centerline{\psfig{figure=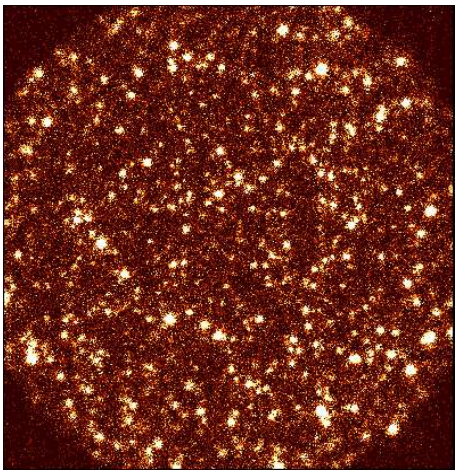,width=6cm,angle=0.0}}
\caption[]{Simulation of a 2-10 keV deep XMM EPIC pn image of 350 ks}
\end{figure}

XMM is the most sensitive X-ray observatory to survey the X-ray sky at photon
energies above 2 keV.  Although its point-spread-function is
significantly worst than that of Chandra, at energies above 2 keV both
instruments are photon-starved and then the much larger collecting
area of XMM will make it more efficient.  We have carried out
extensive simulations of XMM EPIC observations at various depths and
found that the deepest planned XMM observations (PVCal, GTO and AO-1)
reaching ~350-400 ks will {\it not} be confusion noise limited. Figure
1 shows the resulting image in the 2-10 keV band of a simulation of a
350 ks XMM EPIC-pn exposure in a blank field (using the standard model \cite{b23}).  As the EPIC field of view is $\sim 30'$ in diameter,
we expect to find $\sim 300$ sources in such a pointing once
vignetting has been corrected for.  Most of these sources are expected
to lie at redshifts $z>1-2$, from which the X-ray volume emissivity in
hard X-rays will be derived and compared with the one, assumed so far,
obtained with ROSAT for soft X-ray photons.

\begin{figure}[!h]
\centerline{\psfig{figure=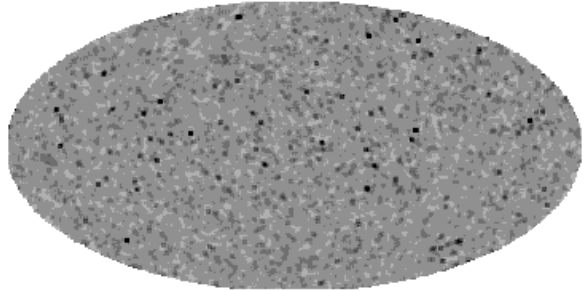,width=8cm,angle=0.}}
\caption[]{Simulation of a 4-12 keV X-ray background map (ignoring the
galaxy), as observed with a 1 $m^2$ collimator with a field of view of
1$\deg^2$, scanning the sky for 6 months 100\% efficiency.}
\end{figure}

\section{X-ray background surveys}

The measurement of large-scale structure in the X-ray
Universe does not necessarily require individually resolving all
sources in very large areas of the sky down to very faint fluxes.  If
the X-ray volume emissivity as a function of $z$ can be derived from
deep surveys, fluctuation analyses of the XRB can also be used
(\cite{b27}). If the X-ray volume emissivity
peaks at some intermediate redshift as it does in soft X-rays ($z\sim
1-2$), then for a fixed angular scale the XRB fluctuations are related
almost uniquely to the value  of the power spectrum of the
inhomogeneities in the Universe at a single comoving wavenumber. The
scales to be probed by XMM (from a few to a few tens of arcmin) will
provide a measurement of the $k\sim (0.1-1)h {\rm Mpc}^{-1}$ regime.
The power spectrum is expected to peak at $k\sim 0.05h {\rm Mpc}^{-1}$, which
corresponds to an angular scale of $\sim 1\deg$.

In \cite{b27} we argue that to detect the excess fluctuations of
the XRB due to source clustering for a beam size of 1$\deg^2$, a
large fraction of the sky needs to be surveyed. To prove that this is
feasible, we have carried out simulations of hard X-ray source
populations over the whole sky with a simple clustering
model for the sources and measured XRB intensities (details in \cite{b27}). These intensities are then `measured' with a proportional
counter of 1m$^2$ effective area during one complete 6-month scan of the sky
at 100\% efficiency and including stable particle background in a
manner similar to the {\it Ginga} LAC observations.

\begin{figure}[!h]
\centerline{\psfig{figure=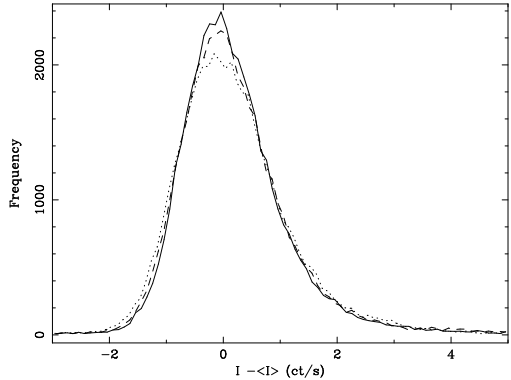,height=5cm,angle=270.}}
\caption[]{Intensity distributions for the map in Fig 2. (ignoring
$\mid b \mid < 20^{\circ}$) for no clustering (continuous line),
linear evolution of clustering (dashed line) and comoving clustering
evolution (dotted line).}
\end{figure}

Figure 2 shows one of these simulations where the clustering has been
modeled with a gaussian correlation function with comoving
evolution. Ignoring data within $\mid b\mid < 20^{\circ}$, Figure 3
shows the histograms for the XRB intensities in 3 cases: absence of
clustering, linear clustering evolution and comoving clustering
evolution ($b_X=1$ in all cases).  The distributions are clearly
distinguishable, and the excess fluctuations can be determined with a
very high accuracy.  Indeed with a significantly smaller collecting
area and similar circumstances, excess fluctuations can still be
detected, but measured with larger statistical uncertainties.

One such survey will also benefit other approaches to measure
large-scale structure in the Universe, particularly the multipole
expansion, especially if a sensitive source survey could also be
carried out. In this way, those $1\deg^2$ regions where sources above
a given flux are detected could be masked out for the multipole
analysis, as discussed in \cite{b12}.

\section{Large-area surveys}

Direct measurements of the large-scale structure of the Universe at
redshifts $z\sim 1-2$ via X-ray observations require surveying areas
of hundreds of square degrees to a sufficient depth. Using galaxy clusters
as tracers of large-scale structure of the Universe presents the
difficulty of the faintness of most of these objects beyond redshifts
$z\sim 1$. Even XMM will require
a large amount of time to do a sensible mapping of galaxy
clusters out to these redshifts.  
 
\begin{figure}[!h]
\centerline{\psfig{figure=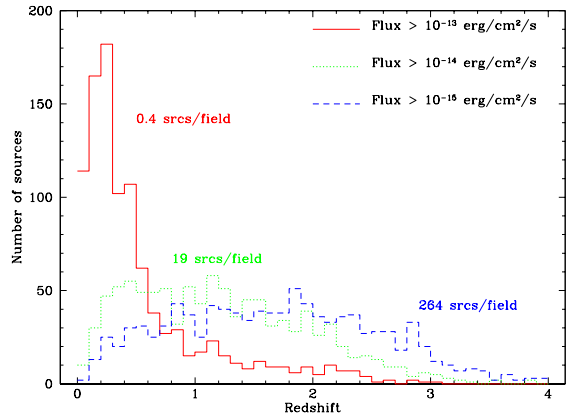,width=8cm,angle=270.}}
\caption[]{Redshift distribution of AGN for different 2-10 keV flux limits.}
\end{figure}

Using AGN has the advantage that they have strong positive evolution
up to $z\sim 1-2$.  In order to reach the redshifts where most of the
X-ray emissivity is produced $z\sim 1-2$, AGN surveys have to go down
to, at least, a 2-10 keV flux $\sim 10^{-14}\, {\rm erg}\, {\rm
cm}^{-2}\, {\rm s}^{-1}$. Figure 4 illustrates the redshift
distribution for different flux limits assuming the \cite{b23} model. The advantage of X-ray AGN surveys over similar
optical work (e.g. the Sloan Digital Sky Survey) is that with hard
X-rays the absorbed AGNs can also be used up to earlier times if
they evolve more strongly than the unabsorbed broad-line objects.

Mapping a 100$\deg^2$ contiguous area of the sky with
the XMM EPIC cameras, which have the largest field of view among all
operating X-ray facilities, to a depth of $\sim 20 ks$ (needed to get
reliable detections at $>10^{-14}\, {\rm erg}\, {\rm
cm}^{-2}\, {\rm s}^{-1}$) will require 10 Msec of effective exposure
time with XMM.  This will collect $\sim 10^5$ sources.

A more efficient way to carry out that project is by means of a dedicated
mission with a wide-field X-ray telescope.  The Panoram-X mission
proposed in \cite{b28}, would cover the whole sky to a
depth $\sim 10^{-14}\, {\rm erg}\, {\rm cm}^{-2}\, {\rm s}^{-1}$.  Of
course, finding redshifts for a reasonable fraction of the several
tens of millions of sources to be discovered by such mission is simply
impossible.  A full analysis of the spatial distribution of X-ray
selected AGN would then require the modelling of the 2D distribution
from these maps, using redshift distributions from the deep hard X-ray
surveys. 

%\begin{figure}[!h]
%\centerline{\psfig{figure=map3.ps,width=6cm,angle=0.}}
%\caption[]{Intensity distributions for the map in Fig 2. (ignoring
%$\mid b \mid < 20^{\circ}$) for no clustering (continuous line),
%linear evolution of clustering (dashed line) and comoving clustering
%evolution (dotted line).}
%\end{figure}

\section{Outlook}

X-ray cosmology is just in its infancy.  Basic questions such as what
is the bias parameter of different classes of X-ray sources are still
partly unanswered.  However the fact that most of the X-ray sky is
dominated by extragalactic sources of which AGN are the major
component make the X-ray sky especially suited for cosmology at
intermediate redshifts.

Making quantitative progress in this field requires not only proper
use of existing or planned observatory-type facilities (i.e., Chandra
and XMM), but probably also dedicated missions to survey all (or most
of) the sky. X-ray cosmology is now in a position to make specific
predictions for the structure of the X-ray universe (once hard X-ray
surveys have been carried out with Chandra and XMM). These surveys can
then be designed and optimized to obtain detections and precise
measurements of the large-scale structure of the Universe at redshifts
$z\sim 1-2$. That would really be a major boost for cosmology at
intermediate redshifts.

%\acknowledgements{We cannot be grateful to the spanish CICYT for
%funding, as project ESP99-0936 has been rejected. (es co\~na)}

%
\vfill
\end{document}